\documentclass[12pt]{article}  
\usepackage{graphicx}

     \newcommand{\hhh}{\hspace*{9mm}}
                            \newcommand{\no}{\noindent}
 \newcommand{\bc}{\begin{center}}
 \newcommand{\ec}{\end{center}}
                   \newcommand{\bfr}{\begin{flushright}}
                   \newcommand{\efr}{\end{flushright}}
   \newcommand{\ii}{\item}
     \newcommand{\be}{\begin{enumerate}}
     \newcommand{\ee}{\end{enumerate}}
        \newcommand{\bi}{\begin{itemize}}
        \newcommand{\ei}{\end{itemize}}
            \newcommand{\bd}{\begin{description}}
            \newcommand{\ed}{\end{description}}
                \newcommand{\beq}{\begin{equation}}
                \newcommand{\eeq}{\end{equation}}
                  \newcommand{\bea}{\begin{eqnarray}}
                  \newcommand{\eea}{\end{eqnarray}}

      \newcommand{\bfi}{\begin{figure}}
      \newcommand{\efi}{\end{figure}}
\newcommand{\bay}{\begin{array}{l}}
\newcommand{\eay}{\end{array}}
            \newcommand{\dd}{\mbox{d}}

    \newcommand{\pa}{\partial}
    \newcommand{\del}{\delta}
    \newcommand{\Del}{\Delta}

    \newcommand{\al}{\alpha}
    \newcommand{\sig}{\sigma}
    
    \newcommand{\Ga}{\Gamma}

\begin{document}
 \pagestyle{empty}
        \hspace*{1mm}  \vspace*{-0mm}
\noindent {\footnotesize {{\em
  }}
\vskip 1.0in
\begin{center}
{\LARGE {\bf Why Fracking Works \\[1.4mm]
             and How to Optimize It
           }}\\[18mm]
{\large {\sc Zden\v ek P. Ba\v zant, Marco Salviato and Viet T. Chau}}
\\[2.1in]

{\large {\sf          Report No. 14-06/008w
} \\[0.8in]

Department of Civil and Environmental Engineering
\\ Northwestern University
\\ Evanston, Illinois 60208, USA
\\[0.6in]  {\bf June 28, 2014}
 }
\end{center}

\clearpage   \pagestyle{plain} \setcounter{page}{1}


\noindent {\bf Abstract:}\, {\sf
      Although spectacular advances in hydraulic fracturing, aka fracking, have taken place and many aspects are well understood by now, the topology, geometry and evolution of the crack system hydraulically produced in the shale still remains an enigma.     
Expert opinions differ widely and fracture mechanicians wonder why fracking works. Fracture mechanics of individual pressurized cracks has recently been clarified but the vital problem of stability of interacting system of hydraulic cracks escaped attention. Progress in this regard would likely allow optimization of fracking and reduction of environmental footprint. The present article first focusses attention on the classical solutions of the critical states of localization instability of a system of cooling or shrinkage cracks in plane strain, and shows that these solutions can easily be transferred to the system of hydraulic cracks. It is concluded that if the profile of hydraulic pressure along the cracks can be made almost uniform, with a steep pressure drop at the front, the localization instability can be avoided. To achieve this kind of profile, which is essential for obtaining crack systems dense enough to allow gas escape from a significant portion of kerogen-filled nanopores, the pumping rate (corrected for the leak rate) must not be too high and must not be increased too fast.
    Subsequently, numerical solutions are presented to show that an idealized  system of circular equidistant vertical cracks propagating from a horizontal borehole behaves similarly. It is pointed out that one important role of proppants, as well as acids that promote creation debris in the new cracks, is that they partially help to limit crack closings and thus localization.
Based on the extremely low permeability of gas shale, one must imagine formation of hierarchical progressively refined crack systems in which the finest cracks have spacing in the sub-centimeter range. Compared to systems of new cracks, the system of preexisting natural cracks is shown to be slightly more prone to localization and thus of no help in producing such a fine crack spacing. The overall conclusion is that what makes fracking work, from the fracture mechanics viewpoint, is the suppression or mitigation of localization instabilities of crack systems, which requires maintaining  
sufficiently uniform pressure profiles along the cracks.
       }

\subsection*{Introduction}

Hydraulic fracturing of oil and gas bearing rocks, aka "fracking"\footnote{
    Although this term, coined in industry, has often been used in the pejorative sense, it is adopted here because of its brevity.},
    is an established technology \cite{Bec10,MonSmi10,SPE10, AdaDet08, GalRee--07} that has been developed gradually since 1947, until recently with no government support. Although the recent advances in fracking have been nothing less than astonishing, the knowledge in the actual fracturing  is mostly empirical and makes a mechanician wonder: Why the fracking works?

The intent of this article is to suggest an explanation in terms of stability of interactive cracks systems. %
However, development of a complete and predictive model is beyond the scope of this study.

\subsection*{An Aper\c cu of Fracking Technology}

The gas bearing stratum of tight shale, typically about 3 km below the surface and 20 m to 150 m in thickness \cite{CipMayWar09,RijCoo01}, is accessed by parallel horizontal boreholes emanating from a single vertical well in the direction of the minimum principal tectonic stress $S_h$, whose magnitude is often about 1/2 to 2/3 of the overburden stress $S_g \cite{HaiHanGuo13}$.

The boreholes are typically about 500 m apart and several kilometers long. Each of them is subdivided into about 5 segments, each of which consists of about 5 to 9 fracturing stages. Each stage, about 70 m long, is further subdivided into about 5 to 8 perforation clusters. In each cluster, about 14 m long, the steel casing, of typical inner diameter 3.5 in. (77 mm) \cite{TanTanWan11,Gal02}, is perforated at 5 to 8 locations by detonating shaped charges
(Fig. \ref{f1}).

Powerful pumps on the surface drill pad inject the fracking fluid into the shale stratum. The fluid, with a proppant (fine sand) mixed into it, is about 99\% water with various additives, such as gellants, acids or pH controlling ions. Each stage requires injection of several million gallons of water (which is equivalent to about 1 to 2 mm of rain over the area of the lease, 3$\times$5 km$^2$). The water that later returns to the surface and represents only about 15\% of the injected total, is highly contaminated. Strict controls are required to prevent its accidental release to the environment. Often, the water outflow is reinjected underground. Avoidance or minimization of this outflow is an important objective of technology improvement.

Pumps, currently attaining at the surface level the pressure of about 25 MPa, force the fracking fluid through perforations in the casing into the shale stratum. The shale is intersected by a system of natural fractures or rock joints, nearly vertical, most of which are filled by calcite or other minerals. They are typically 15 to 50 cm apart \cite{GalRee--07,Ols04}. The shale also contains numerous finer faults, slip planes and near-horizontal bedding planes with millimeter spacing. The first, large, hydraulically produced cracks must obviously be roughly normal to the horizontal wellbore, which is always drilled in the direction of the minimum tectonic stress. These initial cracks may be expected to run preferably along the rock joints, if they have the right orientation.


To provide conduits for gas escape, the crack must branch repeatedly into a hierarchy of secondary and tertiary cracks until a close enough crack spacing is achieved. Most of the branching probably occurs by initiation of secondary cracks from the faces of previous larger cracks at the spacing of dominant inhomogeneities and discontinuities such as rock joints. Branching in a homogeneous material can occur only if the crack is running at nearly the Raleigh wave speed, which is not the case here since the fracking takes several days and thus is quasi-static (sound-generating dynamic crack jumps, of course, do occur, due to inhomogeneities, but long jumps are impossible because the fluid cannot spread so fast). Some static branching at the tip can nevertheless occur at joint intersections, provided that the joints are nearly orthogonal to the crack so that the pressure rise in one branch could not and shield the other, keeping it closed.

\subsection*{A Fracture Mechanics Puzzle}

Most of the gas, principally methane, is contained in isolated kerogen-filled nanopores of diameters from 0.5 nm to about 10 nm \cite{LouReeRup09, JavFisUns07, Jav09}. From drilled cores brought to the surface, the gas content per unit volume of shale is known, and thus it is estimated that about 15\%, and often as little as 5\%, of the gas content of the shale layer gets extracted by the fracking process (estimates as high as 55\% have been heard but are undocumented and doubtful).

Although this percentage seems low, it is nevertheless a puzzle why the percentage is not orders of magnitude lower, given the extremely low permeability of shale. Its value ranges from $10^{-9}$ Darcy to $10^{-7}$ Darcy (which is 10 to 1000-times lower than the typical permeability of concrete) \cite{Mau--Pij-10}. If possible effects of surface forces are ignored, the Darcy law may be written as $v = - (k/\mu) \nabla p$ where $k$ = Darcy permeability and $\mu$ = dynamic viscosity of the gas \cite{AdaDet08,AdaSiePei07}.

As an approximation, one may consider that the pressure profile in length coordinate $\xi$ is fixed and proportional to $(1-\xi/x)^u$ where $u$ = constant (taken here as 2, which is good for linear diffusion), and $x$ = penetration depth of fluid, equal or nearly equal to the crack depth. Also, the pressure gradient at the crack mouth may be approximated as $\nabla p = p/ux$, and the velocity at crack mouth as $v = \dd x /\dd t$ where $t$ = time. Thus the Darcy law leads to the approximate differential equation:
 \beq \label{e1}
        \frac{\dd x}{\dd t} = \frac{ku} \mu \ \frac p x
 \eeq
Solving this equation for constant gas pressure $p$ indicates that the gas penetration depth increases as
 \beq  \label{e2}
  x = \sqrt{ \frac{2ku} \mu \, p\, t }
 \eeq

Substituting the aforementioned minimum and maximum values of shale permeability and assuming a pore pressure of about 25 MPa, one finds that, during 30 days, the pressure front will propagate to the distance $a$ = 0.04 m or 0.44 m, respectively; during one week, $a$ = 0.02 m and 0.22 m. If there is a crack at that distance from the pore, the gas can begin escaping from that pore. Once distance $a$ gets penetrated, the gas can begin draining from the shale pores, and it will take further time for the gas to get extracted. This drainage occurs from both faces of the vertical crack. According to this argument, the shale volume from which the gas can be drained into the vertical crack cannot be grater than the product of $2a$ with the area of the vertical crack.

Further questions arise with regard to the network of preexisting cracks. Many or all of them are cemented by calcite and, whether or not filed, the overburden and tectonic stresses are so high that no gaps or open voids can exist in these cracks. So they can provide no natural conduits for gas escape, unless opened by fracking fluid pressure. But this depends on localization, which is discussed later.

The very small distance $a$ must now be considered from the viewpoint of stability of parallel crack systems \cite[sec.12.5]{BazCed91}. Under many circumstances, parallel cracks tend to localize into one large crack. Consider the localization in one fracturing stage, typically 70 m long, in a shale stratum 150 m deep. If the vertical parallel cracks in this stage are localized into one vertical crack, the gas could be extracted to the depth of 0.04 m or 0.44 m from each face of this crack (Fig. \ref{f2}).

For the aforementioned minimum and maximum permeability values and start at 30 days, one would thus estimate that, respectively, only 0.15\% or 1.47\% of gas contained in the shale layer could be extracted. Yet the industry performance demonstrates that far more gets extracted (up to 15\%). This discrepancy has not yet been explained in terms of fracture mechanics of interactive crack systems. Explanation, understanding and mathematical modeling is a prerequisite for improving and optimizing the fracking process.

\subsection*{Review of Stability of Parallel Crack Systems}

In the mid 1970s, extensive studies of extracting heat from hot dry rock located relatively close to the earth surface were conducted. It was speculated that if a large vertical crack were created hydraulically from a borehole in hot granite and were then intersected by another borehole, circulation of water could deliver enough steam to generate electricity, like in geothermal basins with natural circulation \cite{BazOht78}. Because of rapid decay of the heat conduction flux from a hot wall, success would have required many closely spaced parallel cooling cracks to propagate to a long distance from the walls of the large vertical crack. However, drilling into the giant Valles Caldera in the Jemez Mountains of New Mexico gave a negative result and a study of the localization instability of parallel cooling cracks explained why. Nevertheless, this negative result provides today a valuable lesson for the fracking process.

Consider a system of interacting cracks of lengths $a_1, a_2,...a_N$ in a cooled (or shrinking) solid with fracture energy $\Ga$, and assume applicability of the linear elastic fracture mechanics (LEFM). The Helmholtz free energy has the general form;
 \beq \label{e3}
  {\cal F} = U(a_1, a_2,...a_N; p) + \sum_{i=1}^N \int \Ga \dd a_i
 \eeq
where $U$ is the strain energy of the elastic solid; $p$ is the control parameter, such as the depth of penetration $D$ of the cooling front into the  halfspace (or, in our case considered later, the fluid pressure at the surface of halfspace).
There are many possible fracture equilibrium solutions but thermodynamics requires the system to evolve in such a way that ${\cal F}$ be minimized. The problem is to determine which solution is stable and which solutions are unstable or metastable. The stable solution is what will occur.

The equilibrium and stability of the crack system are decided by the first and second variations \cite{BazOht77}
 \bea \label{e4}
  \delta{\cal F} &=& \sum_{i=1}^m\left(\frac{\partial U}{\partial a_i}+\Gamma \right)\delta a_i+\sum_{j=m+1}^n\left(\frac{\partial U}{\partial a_j} \right)\delta a_j
  \\ \label{e5}
   \delta^2{\cal F} &=& \frac{1}{2}\sum_{i=1}^n\sum_{j=1}^n\left(\frac{\partial^2 U}{\partial a_i \partial a_j} \right)\delta a_i\delta a_j=\frac{1}{2}\sum_{i=1}^n\sum_{j=1}^n {\cal F}_{,ij}\delta a_i\delta a_j
 \eea
(where a unit width $b$ of the crack front is considered); here $i = 1,...m$ are the cracks that are propagating ($\del a_i > 0$), dissipating fracture energy $\Ga$; $i = m+1,...n$ are the cracks that are shortening ($\del a_i > 0$), for which the fracture energy is 0, and $i =  n+1,...N$ are the cracks that are immobile ($\del a_i = 0$), which occurs when the energy release rate $-\pa{\cal U} /\pa a_i$ is non-zero but less than critical.

Equilibrium (or static) crack propagation requires vanishing of the first parenthesized expression in Eq. (\ref{e4}), which represents the Griffith crack propagation criterion of LEFM. There exist many equilibrium solutions, 
reachable along a stable equilibrium path. Fracture stability requires the matrix of ${\cal F}_{,ij}$, equal to $U_{,ij}$, to be positive definite, i.e.,
  \beq \label{p}
   \mbox{det}U_{,ij} > 0~~~\mbox{and}~~~U_{11} > 0
  \eeq
for the vectors of admissible variations $\del a_i$ \cite{BazCed91,BazOht77, BazOhtAoh79, NemKeePar78}.

The admissible crack length variations $\del a_i$ 
are those satisfying the following restrictions:
 \bea \label{u1}
  \mbox{for}~~~~~~\pa U /\pa a_i~~ = \Ga:~~&&\del a_i \ge 0
 \\  \label{u2}
  \mbox{for}~~~0 < \pa U /\pa a_i < \Ga:~~&&\del a_i = 0
 \\ \label{u3}
  \mbox{for}\ ~~~~~~\pa U /\pa a_i~~ = 0:~~&&\del a_i \le 0
 \eea
In the special case of a parallel system of preexisting natural cracks that are open up to length $a_j$ but closed beyond, the effective fracture energy is 0, and then
 \beq  \label{u4}
  \mbox{for}~\pa U /\pa a_j~ =~ 0:~~~\mbox{any}~\del a_j
 \eeq

\subsection*{Localization Instability of Cooling Cracks and Its Analogy with Pressurized Cracks}

Consider a homogeneous elastic halfspace cooled by heat conduction. This  produces a temperature profile in the form of the complementary error function, often approximated by a parabola, whose front advances into the halfspace as $\sqrt{t}$. The thermal stress, proportional to the temperature drop, is considered to produce an advancing system of parallel cooling cracks of equal spacing $s$, whose lengths ideally alternate between $a_1$ and $a_2$. The crack length are assumed to be initially equal, $a_1 = a_2$ (although in reality the crack lengths, as well as spacings, are randomly distributed).

The positive definiteness of the matrix of ${\cal F}_{i,j}$ is first lost by the vanishing of $\det{\cal F}_{,ij}$. But this signifies neither instability nor bifurcation because the corresponding eigenvector ($\del a_1, \del a_2$) implies every other crack to shorten ($\del a_2 = -\del a_1 \ne 0$), which is impossible since the energy release rates $U_{,i}$ (or stress intensity factors $K_i$) of all cracks are nonzero ($K_i = \sqrt{E' U_{,i}}$, $E'$ = elastic modulus for plane strain). An exception is the opening of preexisting (non-cemented, non-sticking) cracks, whose critical energy release rate can be zero, a discussed later.

After further crack growth, when the crack length is about $1.5s$ to $2s$ (depending on ratios $s/D$ and $l_0/D$ where $l_0$ = Irwin's characteristic length), the positive definiteness of the matrix is lost due to the vanishing of ${\cal F}_{,11}$ (and ${\cal F}_{,22}$). The corresponding eigenvector ($\del a_1, \del a_2$) has $\del a_2 = 0$, which is admissible. It implies a stable bifurcation, in which one set of alternating cracks continues to grow ($\del a_1 > 0$), while the remaining cracks get arrested ($\del a_2 = 0$). Later, after further growth of $a_1$, cracks $a_2$ reduce their energy release rate to 0 and close \cite{BazOht78,BazOht77, NemKeePar78}.

The remaining, leading, cracks, with spacing $2s$, eventually reach another bifurcation of the same kind, at which every other crack stops growing and gradually closes while the spacing of open cracks doubles to $4s$ (see Fig. \ref{f3}). This doubling of crack spacing, in which the crack system localizes into fewer and fewer cracks, is periodically repeated as the cooling advances; see \cite[sec. 12.5]{BazCed91}).
Consequently, cooling of the rock by heat conduction cannot reach most of the rock mass (see Fig. \ref{f4}). This fact has rendered the idea of geothermal heat extraction from hot dry rock unworkable

With regard to fracking, it is interesting to recall the 1970s study of the effect of various temperature profiles along the cracks, which could conceivably be altered by heat convection in water flowing along the cracks; see Fig. \ref{f5}, which shows, for several temperature profiles \cite{BazCed91, BazWah79}, the equilibrium curves of crack length versus the cooling front depth. The solid parts of the curves represent stable equilibrium states and the dashed parts unstable equilibrium states. They are separated by the circle points, which indicate bifurcation states at which fracture localizes and every second crack stops growing.

As the temperature profile gets more uniform over a greater portion of crack depth, the bifurcation states are pushed to a greater crack depth. Eventually, for the profiles marked as 4 and 5, which have a long uniform portion and a steep or very steep temperature drop at the end, there is no bifurcation \cite{BazWah79}. So, if such a temperature profile could be produced, the parallel cooling cracks would grow at constant spacing indefinitely.
But for heat extraction from hot dry rock it seems impossible.

\subsection*{
         Equivalence of Localization of Cooling Cracks and Pressurized Cracks}

It is now interesting to point out that the previous analysis of cooling cracks can be easily transferred to fracking, which is a point that has apparently gone unnoticed. To explain, the situation in the left column in Fig. \ref{f6} shows an array of cooling cracks propagating from the surface of a halfspace, opened by temperature drop $\Del T(x)$ which is assumed to depend only on depth coordinate $x$. The thermal stress field is denoted as $\sig_T(x,y)$ (positive for tension). The halfspace is at the same time under tectonic pressure $\sig_h$ in the $y$ direction normal to the surface (negative for compression). The stress intensity factor of the cooling cracks is denoted as $K_I^T$.

The formation of the cooling cracks (left column) can be decomposed into two steps:
 \bd \setlength{\itemsep}{-1.6mm} \ii
I. In the first step (middle column), the cracks are imagined to be glued so as to be kept closed. In that case the temperature field together with the tectonic stress $S_h$ produces, along each cross section $y$ = 0, normal stresses $\sig^T(x) + S_h$ where $\sig^T(x) = - E \al \Del T(x)$ ($\Del T(x) \le 0$); here $\al$ is the thermal expansion coefficient and $E$ is Young's modulus of the rock (considered for simplicity isotropic, although a generalization to orthotropic rock would not be difficult). The temperature drop $\Del T$ is assumed to be big enough for the tensile thermal stress $\sig^T(x)$ to overcome the tectonic stress. Since the cracks do not open, the stress intensity factor in this case vanishes, $K_I$ = 0.
 \ii
II. In the second step (right column), the cracks are imagined to be unglued and allowed to open. This is equivalent to applying onto the crack faces pressure $p(x)= - \sig_T(x)$ that is equal to the stress previously transmitted across the glued cracks ($p$ is positive for compression). The stress intensity factor produced by this pressure is denoted as $K_I^p$. The stress field due to pressure on the cracks is denoted as $\sig_p(x,y)$. The total stress field is $\sig_p(x,y) - S_h$. So,
 \bea  \label{e6}
  p(x)  &=& - E \al \Del T(x)~~~(p(x) \ge 0~\mbox{assumed})
 \\ \label{e7}
  K_I^p &=& K_I^T
 \eea
The singular stress field of the cooling cracks is thus decomposed as $\sig_T (x,y) = \sig_T(x) + \sig_p(x,y)$, and so the singular stress field due to pressuring the unglued cracks may be expressed as
 \beq  \label{e8}
  \sig_p(x,y) = \sig_T(x,y) - E \al \Del T(x)
 \eeq
onto which the tectonic stress $-S_h$ gets superposed (see Fig. \ref{f6}).
 \ed

The foregoing hydro-thermal equivalence of thermal and pressurized cracks can be extended to crack systems of different topology and geometry, e.g., with curved and variously inclined cracks. On the other hand, applicability is limited to the case of LEFM, in which the fracture process zone (FPZ) is assumed to be a point. In quasi-brittle fracture mechanics, the foregoing hydro-thermal equivalence is only approximate. The reason is that the FPZ, considered to have a finite size, is affected by the nonsingular part of stress field, which is different in the left and right columns of Fig. \ref{f6}.

The foregoing studies have been conducted without specifically considering that the cracks preferentially grow along the plentiful natural cracks or joints. Although their detailed consideration will require numerical simulation, qualitatively the same localization behavior must be expected. Closed or filled naturally cemented cracks do not change significantly the stiffness characteristics of the shale mass. When a crack propagates along a weak, naturally cemented, preexisting crack or joint, the only significant difference is that the fracture energy is smaller, perhaps even zero, as discussed later. But this does not change our conclusions about localization qualitatively.

So we may conclude that the effect of temperature profile on fracture propagation is generally the same as that of a similar crack pressure profile. Thus the previous analysis of cooling cracks makes it possible to state, even without any calculations, that by achieving a sufficiently uniform crack pressure profile, with a sufficiently steep pressure front, the parallel cracks should not get localized and should propagate indefinitely, at constant spacing. This is what is needed to create densely spaced channels by which the shale gas could escape from the nanopores. And since the fracking actually works, we must conclude that such pressure profiles are indeed being achieved. It is intuitive that it is mainly a matter of pumping rate. If the pressure at crack mouth were increased in small steps and after each step the pressure was held constant long enough, the pressure in the cracks would eventually become uniform (however, since there is extensive leaking of the fracking fluid into pores and voids other than the cracks, the pumping rate must be corrected for the leaking and what matters is the rate of fluid influx at the crack mouths).

\subsection*{Numerical Results on Localization of Fluid Pressurized Cracks}

The foregoing analysis applies to the plane strain situation, which is a reasonable approximation for vertical cracks spreading wide from the horizontal borehole and over the full depth of the stratum. In an earlier stage (though not right at the start of cracks from the casing perforated in one direction only), these cracks  are probably better approximated as circular cracks. Then the problem is approximately axisymmetric.

To examine the broader validity of the foregoing inferences from thermal stress analogy in plane strain situation, axisymmetric finite elements are now used to analyze the stability of a system of primary vertical circular cracks of equal spacing $s$ (Fig. \ref{f7}), normal to the direction of perforated horizontal borehole. This is, of course, a simple idealization of cracks which are surely quite irregular and propagate preferentially along preexisting cemented joints. The response is assumed to be symmetric with respect to each crack plane, which is again an idealized situation obtained for a crack system infinite in the direction normal to the cracks. For numerical reasons, the body containing the cracks is assumed to be an infinite cylinder with the borehole in the axis. Then it is possible to exploit symmetry with respect to the crack planes and analyze only a slice of the cylinder between two crack planes. The cylinder radius $R$ = 60 s, is considered sufficiently larger than the crack radius.

The tectonic minimum principal stress, which is normal to the cracks, is considered to be $\sig_h$ = 40 MPa. 
The shale is simplified as isotropic, with Young's modulus $E$ = 37,600 MPa, Poisson ratio $\nu$ = 0.3 and fracture energy $G_f$ = 208 J/m$^2$. It is assumed that crack radii alternate as $a_1, a_2$ as shown in Fig. \ref{f7}, and that initially $a_1 = a_2$. The pressure, $p_0$, of the fluid in the borehole is gradually raised and various pressure profiles along the radius are assumed to be maintained the same during the crack growth, given by the equation
 \beq \label{profile}
  \Del p(x') = \left(1 - \frac{x'} D \right)^u (P_0 - \sig_h)
 \eeq

From the results of the bifurcation analysis, plotted in Fig. \ref{f7}, two observations can be made:
\\ \hhh  1) If the stability of the dense crack system is not unlimited, an increase of the pressure applied from the borehole tends to increase the critical crack length at bifurcation and thus tends to stabilize the distributed crack system, i.e., prevent crack closing and localization into one crack.
\\ \hhh  2) For the pressure profiles with a mild pressure decrease over the crack length and a steep pressure drop near the crack tip ($u$ = 1/2 or 3/4, top of Fig. \ref{f7}), the crack system path exhibits no bifurcation in the $(a_1, a_2)$ space and maintains the original crack spacing and equal crack length ($a_1 = a_2$) indefinitely, without localization.



Aside from favorable crack pressure profiles, addition of proppants (fine sand) and acids to the cracking water are other empirically introduced measures that have been proven to help fracking. It is generally considered that their purpose is to keep the cracks open during gas extraction. But not only that. The present analysis shows that the proppants are also important for preventing or partially limiting crack localizations during the hydraulic fracturing process.

Why the acids have been found to help gas extraction is also unclear from the mechanics viewpoint. It is speculated that the acids help to loosen asperities from the crack walls, thus creating fragments and debris that tend to keep the cracks open during gas extraction. But again, like the proppants, this not the only role of acids. Their another role may be partial prevention of crack localizations.

For a fixed profile, the crack system bifurcation problem involves 5 variables: $E, \Ga, p, s, a_1$. Since they involve only 2 independent dimensions, length and force, the solution must depend (according to the Vashy-Buckingham theorem of dimensional analysis) on only 3 dimensionless parameters. They are:
 \beq \label{dimless}
  \Pi_1 = \frac p E,~~~\Pi_2 = \frac{a_1} s,~~~
  \Pi_3 = \frac{p\, s}{\Ga}
 \eeq
In the cooling problem, there is always one more parameter $\Pi_4 = D/s$ where $D$ is the cooling front depth; and here, too, necessitating parameter $\Pi_4 = x/a_1$). Mapping of all possible solutions in terms of these parameters is relegated to a separate study. For dense cracks of small enough spacing $s$ (and also for a larger scale model with smeared joints and flaws), the tensile strength of shale, $f_t$, must also matter, in the sense of the cohesive crack model, and then there is an additional dimensionless parameter $\Pi_5 = l_0 /s$ where $l_0 = E \Ga /f_t^2$ = Irwin's characteristic length for cohesive (or quasibrittle) fracture.

\subsection*{Localization in Preexisting System of Natural Cracks or Rock Joints}

The shale mass typically contains one or few systems of nearly parallel  preexisting natural cracks or joints. Their typical spacing ranges from 0.1 m to 1 m. Due to tectonic and overburden pressures, the opposite faces are in perfect contact and so these opposite cracks and joints cannot  serve as conduits for fluid unless opened up by high enough pressure of the fracking fluid.

Often these natural cracks are filled and cemented by calcite or other minerals. So, their opening requires a finite fracture energy $\Ga$, which may be expected to be smaller than the $\Ga$ of the intact shale. Then, if they are normal or nearly normal to the minimum principal stress, the fracking fluid will open them first. Their behavior, including localization, is similar to new cracks in intact rock.

Some natural cracks might not be cemented by a fill, in which case their fracture energy $\Ga = 0$. Does that make such natural crack system more likely, or less likely, to serve as conduits for extracting gas?---Little less likely, because the natural cracks, while easier to open, are slightly more prone to localization.

The bifurcation state that indicates localization is determined by the vanishing of the second variation $\del^2 {\cal F}$ (Eq. \ref{e4}), which is independent of whether $\Ga$ is finite or zero. However, there is a difference in the admissibility of the eigenvector through which the matrix of ${\cal F}_{,ij}$ ($i,j, = 1,2$) loses positive definiteness (Eq. \ref{p}, ${\cal F}_{,ij} = U_{,ij}$). As pointed out before, the singularity occurs first through the vanishing of ${\cal F}_{,ij}$. The corresponding eigenvector $(\del a_1, \del a_2) \propto (1, -1)$ \cite{BazCed91} which is, in view of Eqs. (\ref{u1})--(\ref{u3}), inadmissible for new cracks because their energy release rate $-U_{,i}$ is non-zero.

However, for un-cemented natural cracks such an eigenvector is, according to Eq. (\ref{u4}), admissible, i.e., every other crack can start shortening as the others extend (the closed crack portion is not counted into the crack length). So for natural cracks, for which $\Ga$ = 0, the localization of parallel cracks will occur earlier in the fracking process than it will for parallel cracks with $\Ga >0$ in intact rock, in which those cracks localize slightly later, when ${\cal F}_{,11}$ vanishes (Eq. \ref{p}), with $(\del a_1, \del a_2) \propto (1, 0)$.

 \subsection*{Hierarchical Refinement of Hydraulic Crack System}

A possible idealized picture of crack system development may now be offered. From the horizontal borehole, the first vertical cracks, orthogonal to the borehole, must form at the locations of casing perforations. In reality, these primary cracks are sure not to be exactly planar nor exactly vertical, nor exactly circular, since they should preferably follow the irregular near-vertical rock joints.

Then, in the direction normal to the larger principal tectonic stress $\sig_H$, one can imagine that a system of secondary vertical cracks of denser spacing, roughly orthogonal to the primary vertical cracks, would have to form, preferentially following the rock joints or slip faults of roughly that direction. These cracks would likely start by fluid penetration into the rock inhomogeneities such as faults, joints and preexisting cracks, driven by tensile stress parallel to the to the primary crack walls produced by expansion of the fracking zone under fluid pressure.
If a nearly uniform pressure profile with a steep front can be maintained in these secondary cracks, they are likely not to localize and thus maintain their narrow spacing.

From the walls of the secondary cracks, a tertiary system of parallel cracks of still denser spacing must initiate, etc. (see Fig. \ref{f8}). Nearly horizontal parallel cracks with millimeter range spacing might propagate from the vertical cracks along the bedding planes in shale, which represent planes of the lowest fracture energy. The bedding planes may concentrate organic matter containing gas \cite{LouReeRup09}.
Opening of cracks along the bedding planes is very desirable but possible only if the pressure of pumped fluid exceeds the vertical overburden pressure $\sig_g$, which is today only rarely achieved with the currently available pumps.

It is thus clear that, to explain the known percentage of gas extraction, a hierarchical multi-level crack system that leads to fine cracks with millimeter-range spacing must get formed. Since the initiation of cracks from a smooth surface is governed, according to the cohesive crack model, by the tensile strength rather than the fracture energy, tensile stresses parallel to the walls of the higher-level crack must develop. These tensile stresses must be generated as a reaction to the pressurization of a large enough volume of the fracturing zone in shale. In similarity to what is known for concrete, the initial spacing of the sub-level cracks produced tensile stress along the wall of a higher-level crack is expected to be roughly equal to the spacing of shale inhomogeneities, which is also the spacing of weak spots on the wall.

\subsection*{The Intriguing Prospect of Shale Comminution by Shock Waves}

Recent years have seen a revival of the idea to stimulate gas release from shale by shock waves emitted by explosions in the horizontal borehole. Oil companies have been experimenting with various types of explosions in the borehole, including electric pulsed arc \cite{Che--Pij-12, SafHuaMut14}. If shale comminution (i.e., fragmentation, pulverization and crushing) could be achieved over a large zone of the shale, it would allow a full or partial replacement of hydraulic fracturing, which would significantly reduce the outflow of contaminated water.

The shock waves would comminute the compressed shale by high rate shear. To simulate it computationally, a theory inspired by analogy with turbulence has recently been conceived \cite{BazCan13, BazCan14}. At high shear rates, the driving force of comminution under triaxial compression is not the release of strain energy, as in classical fracture mechanics, but the release of the local kinetic energy of shear strain rate of particles being formed by interface fracture. It transpired that at shear strain rates $>10^4$/s, the local kinetic energy density exceeds the maximum possible strain energy density (i.e., the density at the strength limit) by several orders of magnitude.

Since the formulation involves quantities with both the stress or energy density (dimension J/m$^3$) and surface energy (dimension J/m$^2$), there must exist a material characteristic length which governs the particle size (note that if particle breakage were attributed to exceeding the strength limit, the particle size would be infinite). It is found that the particle size or crack spacing should be proportional to the -2/3 power of the shear strain rate, and that the comminution process is macroscopically equivalent to an apparent shear viscosity proportional to the -1/3  power of the shear strain rate. A dimensionless indicator of the comminution intensity is formulated. The theory was inspired by noting that the local kinetic energy of shear strain rate plays a role analogous to the local kinetic energy of eddies in turbulent flow.

\subsection*{Key Points and Conclusions}
   \be \setlength{\itemsep}{-1.5mm} \item
What makes fracking work is the prevention or mitigation of the localization instabilities of parallel crack systems.
 \ii 
Based on the extremely low Darcy permeability of shale, extraction of a significant percentage of gas requires comminuting the shale stratum by cracks of sub-centimeter spacing.
 \ii
While the preexisting uncemented (unfilled) natural cracks or joints are easier to open, they are more prone to localization and thus do not help in achieving the afore-mentioned fine comminution.
 \ii
In LEFM, the problems of pressurized cracks and cooling cracks are analogous. Whether or not a hydraulic crack system would localize into sparse wide cracks can be easily inferred from the previous studies of cooling cracks (or shrinkage cracks).
 \ii
The key to preventing localization is to achieve and maintain a sufficiently uniform pressure profile, with a sufficiently steep pressure drop at front. This calls for sufficiently slow rise of pressure at the point of fluid injection from the perforated casing into the shale mass, which is controlled by the pumping rate corrected for leaks.
 \ii
Aside from the pressure profile, what also helps is to prevent crack  localization instabilities is the proppant, as well as the fragments created by loosening of asperities from crack walls (which may be promoted by acids in the fracking fluid). However, the proppant or fragments can be only partially effective against localizations since they cannot prevent partial closing of cracks wider than the grain or fragment size.
 \ii
Computer simulation of hypothetical crack evolution and pressure profiles based on pumping history may be expected to help to optimize the fracking process.
 \ee

\no {\bf Acknowledgment:} Funding from the U.S. Department of Energy through subcontract No. 37008 of Northwestern University with Los Alamos National Laboratory is gratefully acknowledged. Crucial initial funding was provided under Grant 36126 by Institute for Sustainable Energy (ISEN) of Northwestern University. Thanks for valuable comments are due to Professor Charles Dowding of Northwestern University and to Norm Warpinski, Technology Fellow at
Pinnacle - A Halliburton Service, Houston, Texas.

\listoffigures  
\clearpage

\bfi \center
\includegraphics[clip=true,width = 1.0\textwidth]{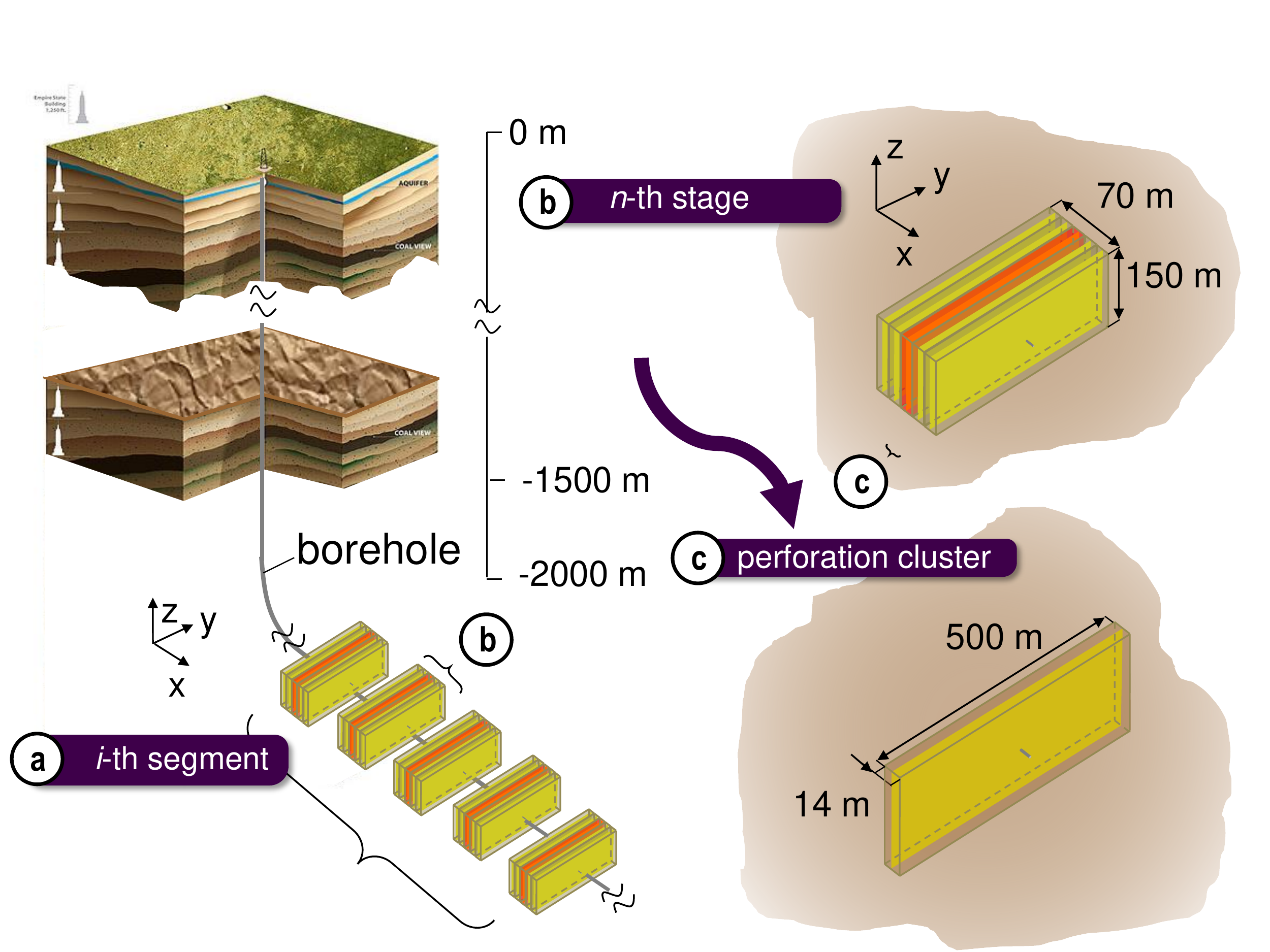} \caption{\label{f1} \sf Overall scheme of hydraulic fracturing: a) Segment subdivided in 5-8 fracturing stages; b) Fracturing stage composed of 5-8 pipe perforation cluster; c) a perforation cluster with 5-8 perforation along the pipes (the figure is not to scale).} \efi
\clearpage

\bfi \center
\includegraphics[clip=true,width = 1.0\textwidth]{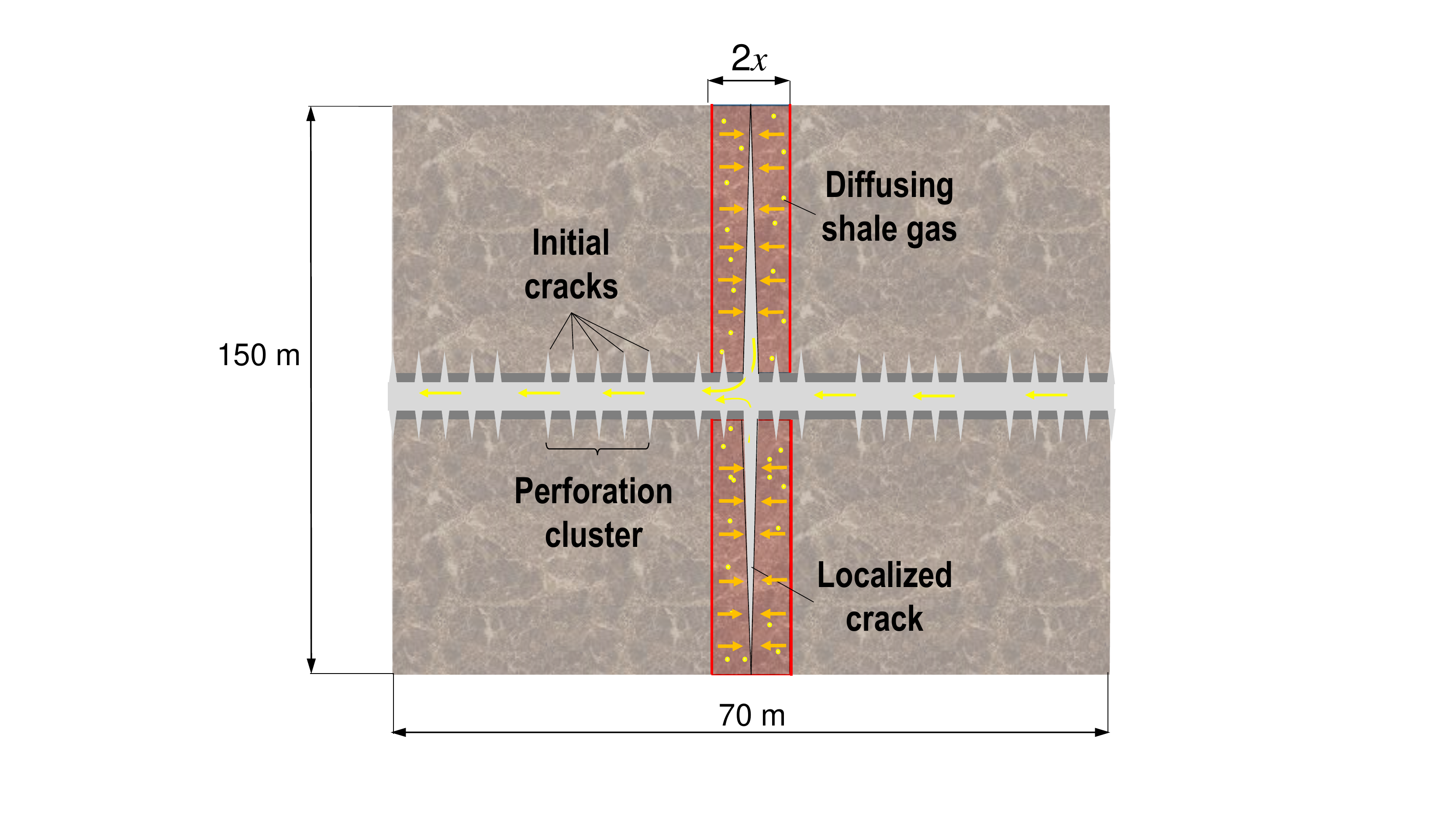} \caption{\label{f2} \sf Schematic of a fracturing stage considered for estimating the amount of gas extracted for the case of one localized crack (for the sake of simplicity, the cracks are assumed to be approximately circular, as a crude approximation; the figure is not to scale. } \efi
\clearpage
\bfi  \center
\includegraphics[trim=6cm 1cm 6cm 1cm,clip=true,width = 1.0\textwidth]{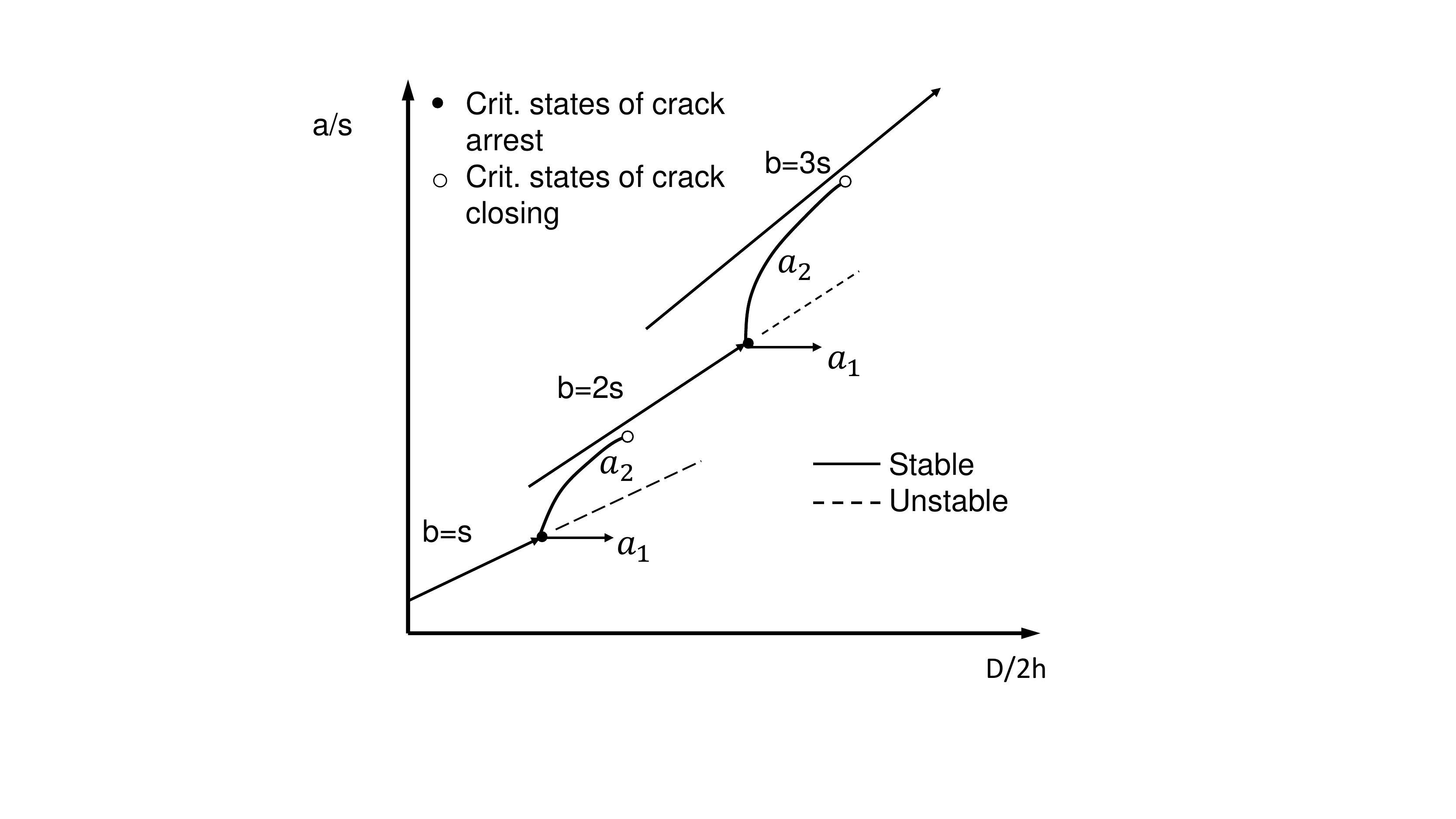} \caption{\label{f3} \sf Path of the lengths of thermal cooling cracks in the crack length space (adapted from \cite{BazOht78}).} \efi
\clearpage
\bfi \center
\includegraphics[trim=6cm 1cm 6cm 1cm,clip=true,width = 1.0\textwidth]{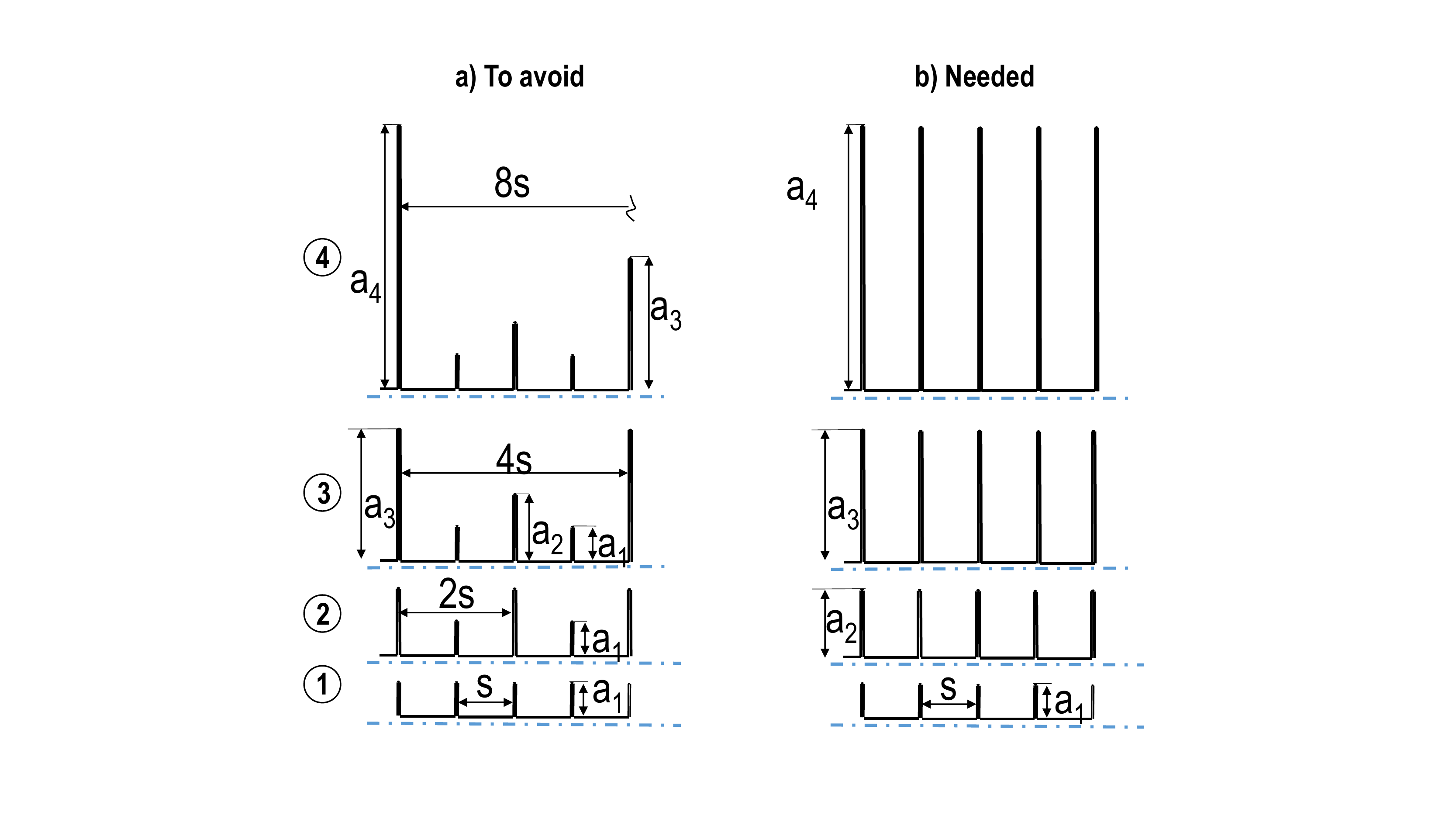} \caption{\label{f4} \sf Schematic representation of fracturing behavior a) with crack localization---undesirable, and b) without crack localization---desirable (prevention of localization increases the percentage  of gas that can be reached from the shale stratum by hydraulic fracturing). } \efi
\clearpage

\bfi \center
\includegraphics[trim=3cm 0cm 3cm 0cm,clip=true,width = 1.0\textwidth]{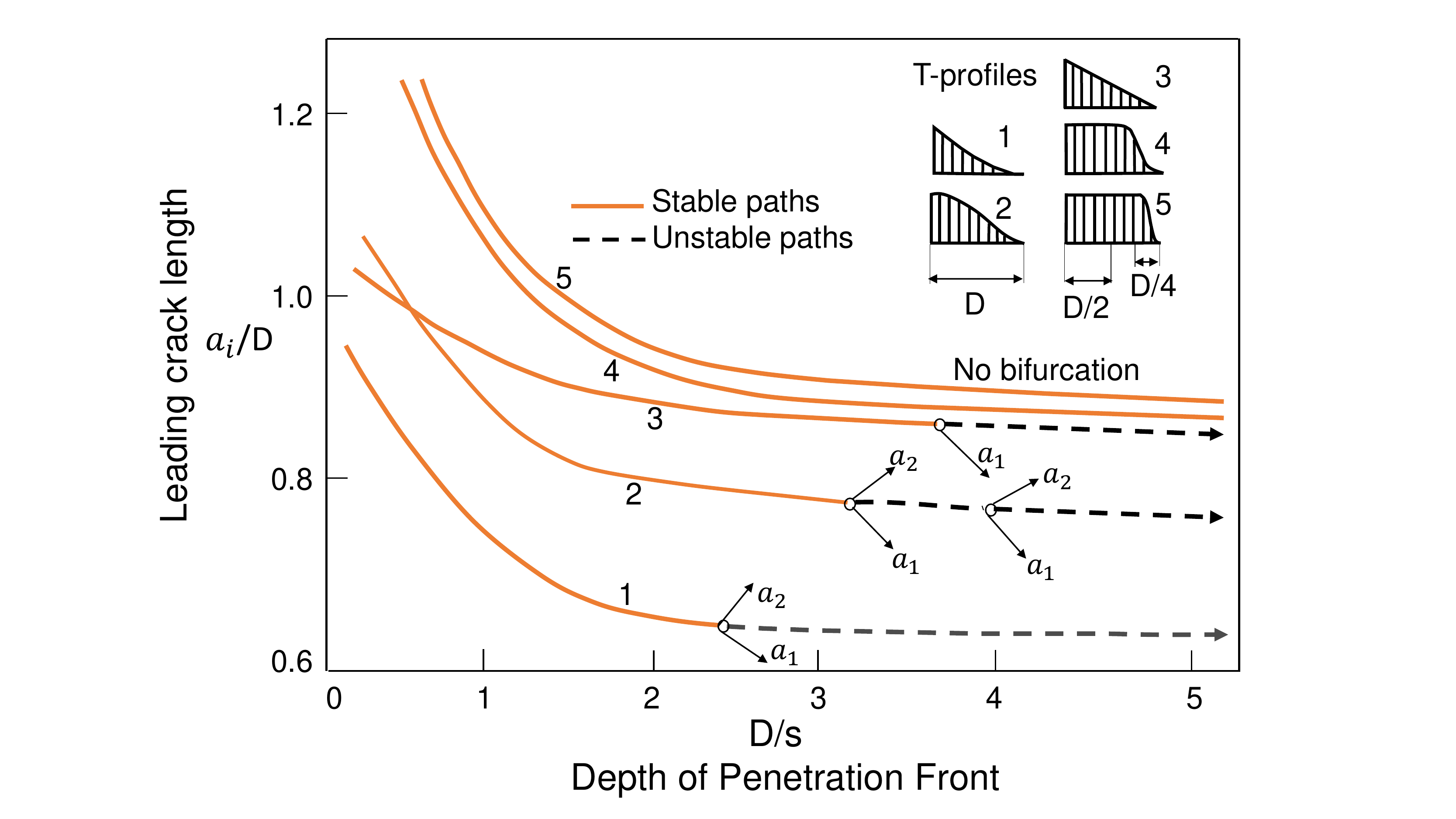} \caption{\label{f5} \sf Leading crack length as a function of the depth of penetration front, for different temperature profiles along the cracks (adapted from \cite{BazOht78}). } \efi
\clearpage

\bfi \center
\includegraphics[trim=3cm 0cm 3cm 0cm,clip=true,width = 1.0\textwidth]{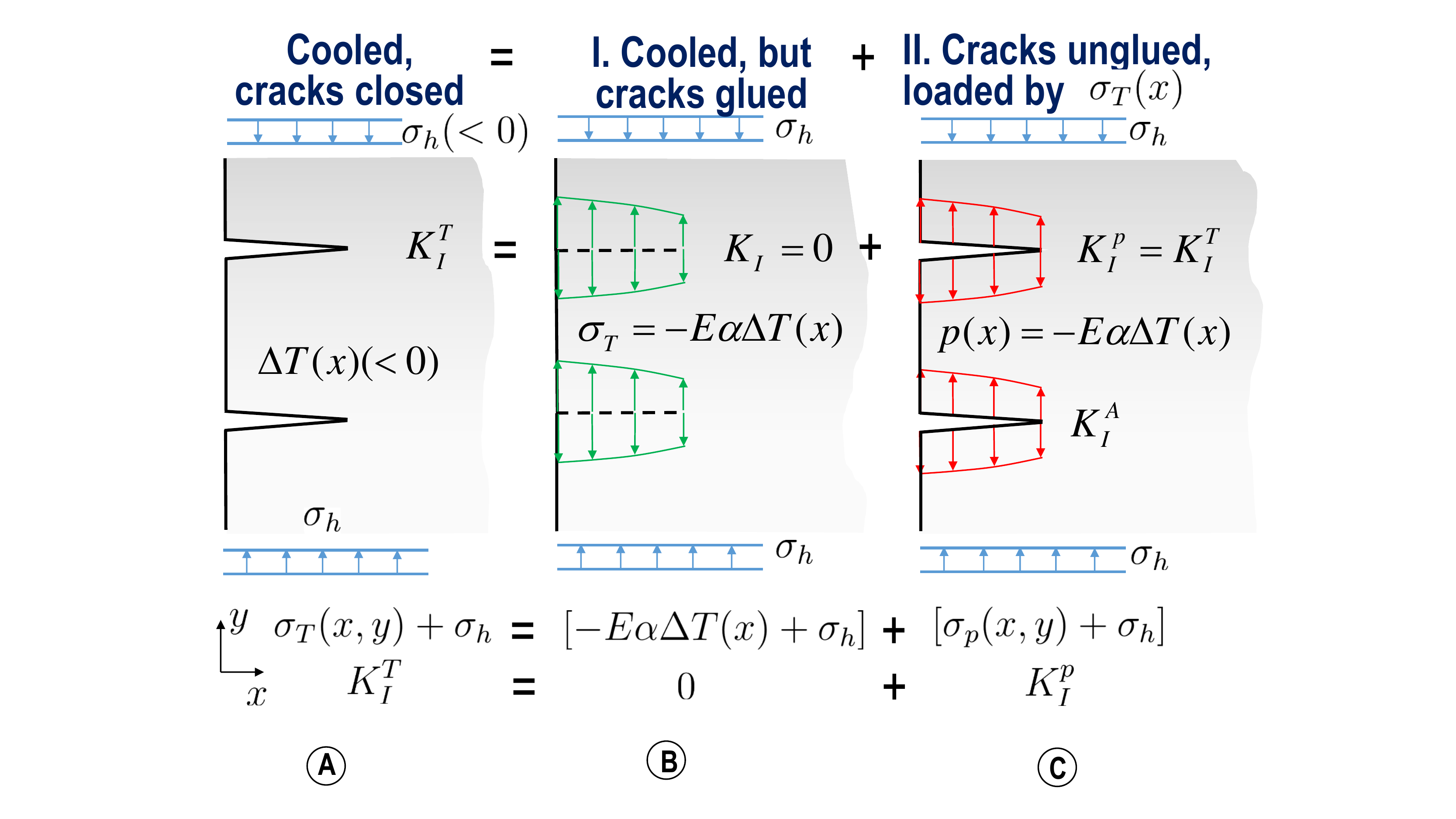} \caption{\label{f6} \sf Analogy of thermal and hydraulic cracks. The formation of the cooling cracks (A) can be decomposed into two steps: In the first step (B) the cracks are imagined to be glued so as to be kept closed. In the second step (C), the cracks are imagined to be unglued and allowed to open. } \efi
\clearpage
\bfi \center
\includegraphics[trim=3.5cm 8cm 3.5cm 2cm,clip=true,width = 1.0\textwidth]{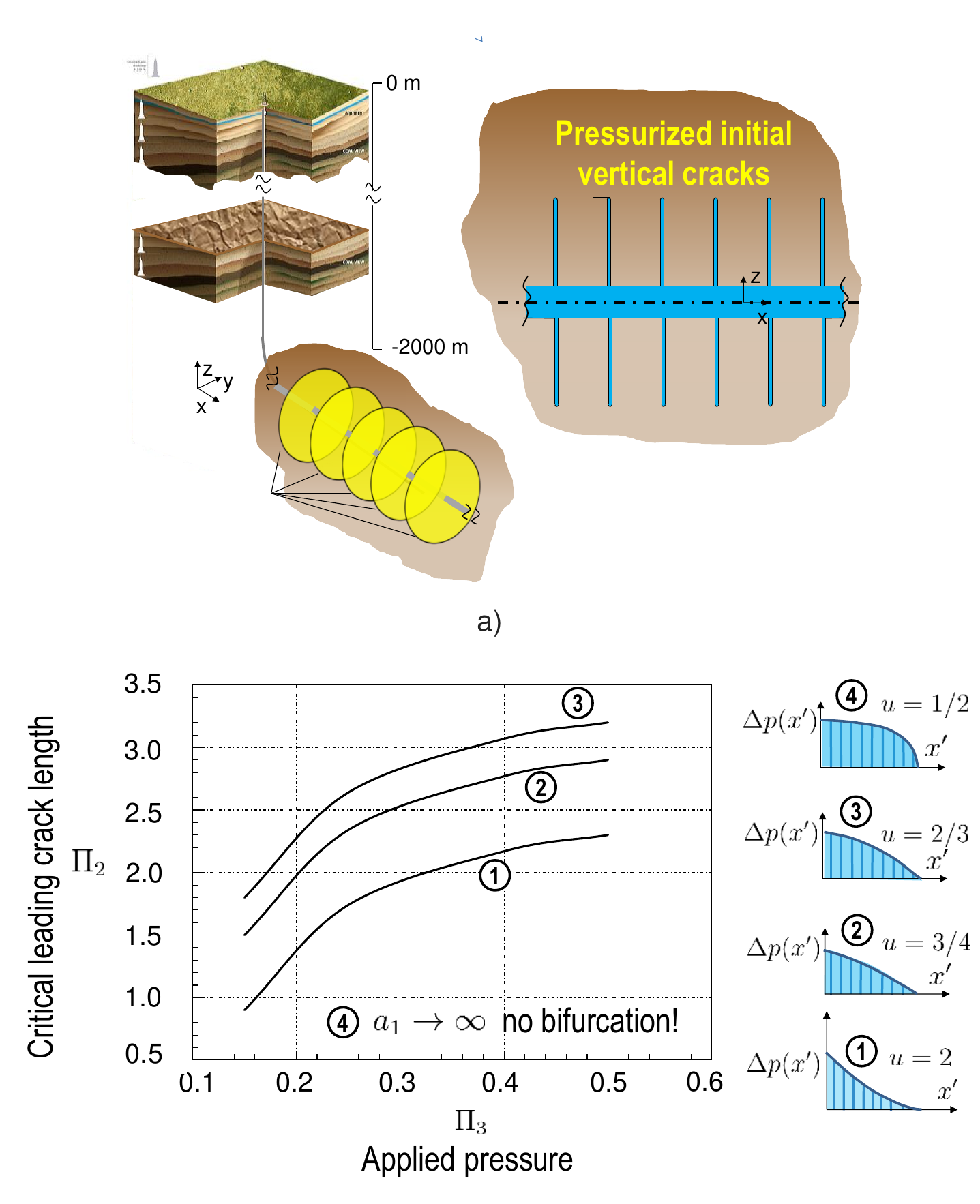} \caption{\label{f7}a) \sf  Idealized circular hydraulic cracks around a horizontal borehole considered for simple analysis of localization instability (not to scale); and b) critical crack lengths as a function of applied pressure for different hydraulic pressure profiles shown. The results show that nearly uniform pressure profiles prevent localization. } \efi
\clearpage
\bfi \center
\includegraphics[trim=6cm 3cm 6cm 1cm, clip=true,width = 1.0\textwidth]{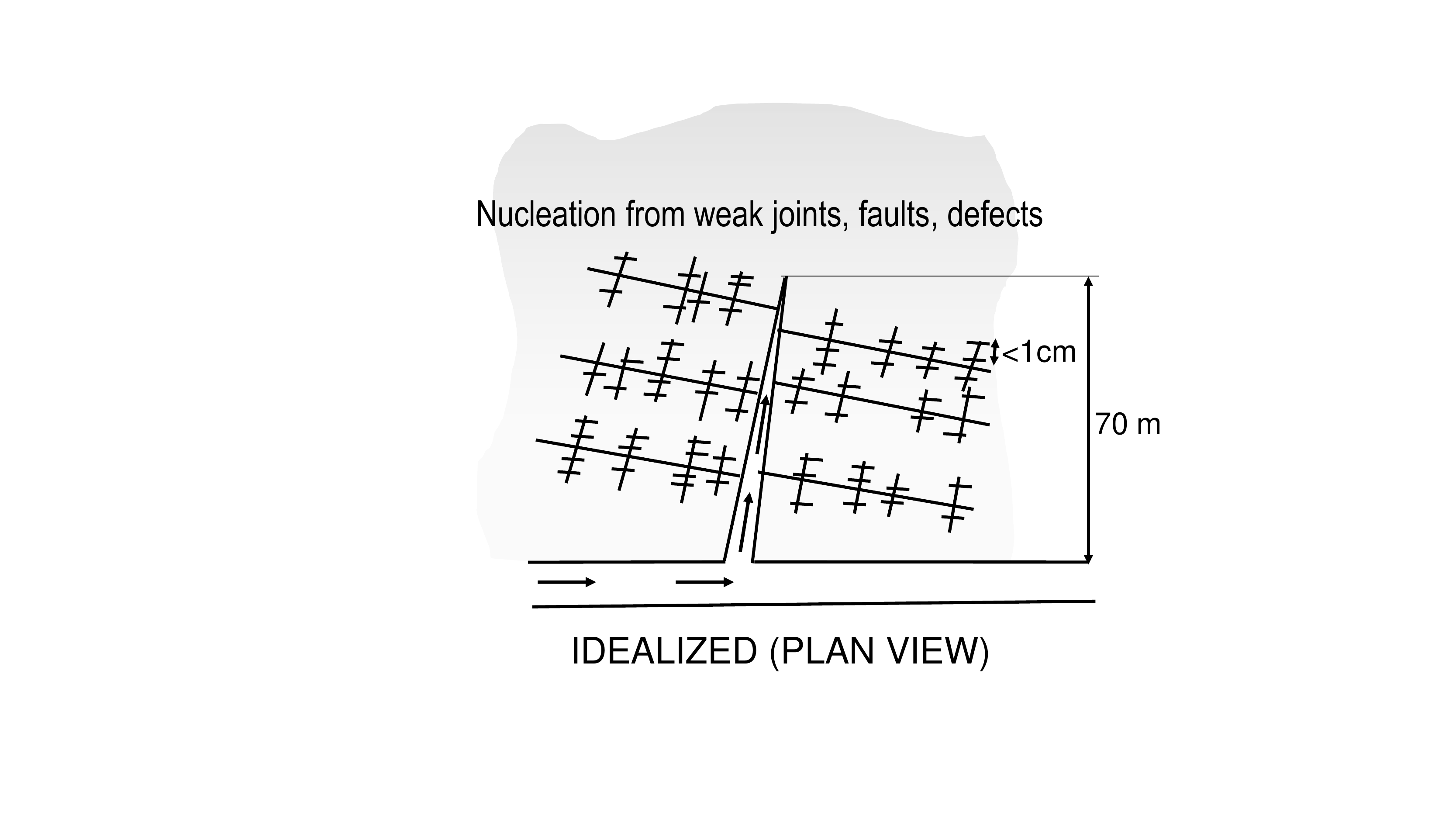} \caption{\label{f8} \sf Schematic picture of how a hierarchical system of hydraulic cracks can lead to crack spacing $<$1 cm. The first nearly vertical cracks form at the locations of casing perforations, preferentially along near-vertical rock joints. Second, in the direction nearly normal to the larger principal tectonic stress $\sig_H$, secondary vertical cracks of denser spacing form, preferentially following secondary rock joints or slip faults. Third, a tertiary system of nearly parallel cracks of still denser spacing propagates roughly orthogonally from the faces of secondary cracks; etc. Near-horizontal cracks along bedding planes can also form if the fluid pressure exceeds overburden pressure (crack branching at tips unlikely). } \efi


\begin{thebibliography}{99}\setlength{\itemsep}{-2mm}
     {\small  


\bibitem{Bec10} Beckwith, R. 2010, ``Hydraulic fracturing: The fuss, the facts, the future" {\em J Petrol Tech}, \textbf{63} (12), 34--41.

\bibitem{MonSmi10} Montgomery, C.T., and Smith, M.B. 2010, ``Hydraulic fracturing: History of an enduring technology", {\em J Petrol Tech}, \textbf{63} (12), 26--32.

\bibitem{SPE10} Society of Petroleum Engineers 2010, ``Legends of Hydraulic Fracturing" (CDROM), ISBN:978-1-55563-298-4

\bibitem{AdaDet08}  Adachi, J.I., Detournay, E. 2008, ``Plane strain propagation of a hydraulic fracture in a permeable rock." {\em Eng Fract Mech} \textbf{75}, 4666-4694.

\bibitem{GalRee--07} Gale, J.F.W., Reed, R.M., and Holder, J. 2007, ``Natural fractures in the Barnett shale and their importance for fracture treatments." {\em Am Assoc Petr Geol B} \textbf{91} (4), 603--622.


\bibitem{CipMayWar09} Cipolla, C.L., Mayerhofer, M.J., Warpinski, N.R. 2009, ``Fracture Design Considerations in Horizontal Wells Drilled in Unconventional Gas Reservoirs." {\em SPE Hydraulic Fracturing Technology Conference, 19-21 January, Texas}.

\bibitem{RijCoo01} Rijken, P., and Cooke, M.L. 2001, ``Role of shale thickness on vertical connectivity of fractures: application of crack-bridging theory to the Austin Chalk, Texas." {\em Tectonophysics}, \textbf{337}: 117-33.
\bibitem{HaiHanGuo13} Haifeng, Z., Hang, L. Guohua C., Yawei, L., Jun, S., and Peng, R. 2013, ``New insight into mechanisms of fracture network generation in shale gas reservoir." {\em J Petrol Sci Eng}, \textbf{110}: 193-98.
\bibitem{TanTanWan11} Tang, Y., Tang, X., and Wang, G.Y. 2011, ``Summary of hydraulic fracturing technology in shale gas development." {\em Geol. Bull. China}, \textbf{30}: 393-99.

\bibitem{Gal02} Gale, J.F.W. 2002, ``Specifying lengths of horizontal wells in fractured reservoirs." {\em SPE Reserv Eval Eng}, 78600: 266-72.

\bibitem{Ols04} Olson, J.E. 2004, ``Predicting fracture swarms —The influence of subcritical crack growth and the crack-tip process zone on joint spacing in rock." {\em J Geol Soc London}, \textbf{231}: 73-87.

\bibitem{LouReeRup09} Louck, R.G., Reed, R.M., Ruppel, S.C., and Jarvie, D.M. 2009, ``Morphology, Genesis, and Distribution of Nanometer-scale Pores in Siliceous Mudstones of the Mississippian Barnett Shale." {\em J Sediment Res}, \textbf{79}: 848-861.

\bibitem{JavFisUns07} Javadpour, F., Fisher, D., Unsworth, M. 2007, ``Nanoscale Gas flow in Shale Gas Sediments." {\em J Can Petrol Technol}, \textbf{46}, 55-61.

\bibitem{Jav09} Javadpour, F. 2009, ``Nanopores and Apparent permeability of Gas Flow in Mudrocks (Shales and Siltstones)." {\em J Can Petrol Technol}, \textbf{48}: 16-21.

\bibitem{Mau--Pij-10} Maurel, O., Reess, T., Matallah, M., de Ferron,A., Chen, W., La Borderie, C., Pijaudier-Cabot, G., Jacques, A.,  Rey-Bethbeder, F. 2010, "Electrohydraulic shock wave generation as a means to increase intrinsic permeability of mortar'',  {\em Cement Concrete Res} \textbf{40}, 1631-1638.

\bibitem{AdaSiePei07} Adachi, J., Siebrits, E., Peirce, A. 2007, ``Computer Simulation of Hydraulics Fractures." {\em Am. Int J Rock Mech Min}, \textbf{44}: 739-57.

\bibitem{BazCed91} Ba\v zant, Z.P., and Cedolin, L. 1991, {\em Stability of Structures: Elastic, Inelastic, Fracture and Damage Theories}, Oxford University Press, New York (also 2nd. ed. Dover Publ. 2003, 3rd ed. World Scientific 2010).

\bibitem{BazOht78} Ba\v zant, Z.P., and Ohtsubo, R. 1978, ``Geothermal heat extraction by water circulation through a large crack in dry hot rock mass." {\em Int J Numer Anal Met}, \textbf{2}, 317--327.

\bibitem{BazOht77} Ba\v zant, Z.P., and Ohtsubo, H. 1977, ``Stability conditions for propagation of a system of cracks in a brittle solid." {\em Mech Res Commun}, \textbf{4} (5), 353--366.

\bibitem{BazOhtAoh79}  Ba\v zant, Z.P., Ohtsubo, R., and Aoh, K. 1979, ``Stability and post-critical growth of a system of cooling and shrinkage cracks." {\em Int J Fracture}, \textbf{15}, 443--456.

\bibitem{NemKeePar78} Nemat-Nasser, S., Keer, L.M., and Parihar, K.S. 1976, ``Unstable growth of thermally induced interacting cracks in Brittle solids." {\em Int J Solids Struct} \textbf{14}, 409-430.

\bibitem{BazWah79} Ba\v zant, Z.P., and Wahab, A. B. 1979, ``Instability and spacing of cooling or shrinkage cracks." {\em J Eng Mech-ASCE}, \textbf{105}, 873--889.

\bibitem{Che--Pij-12} Chen, W., Maurel, O., Reess, T. de Ferron, A., La Borderie, C., Pijaudier-Cabot, G., Rey-Bethbeder,  F., Jacques, A. 2012, ``Experimental study on an alternative oil stimulation technique for tight gas reservoirs based on dynamic shock waves generated by pulsed arc electrohydraulic discharges." {\em J Petrol Sci Eng} 88-89, 67-74.

\bibitem{SafHuaMut14} Safari, R., Huang, J., and Mutlu, U. 2014, ``3D Analysis and Engineering design of Pulsed Fracturing in Shale Gas Reservoirs." {\em Am. Rock Mechanics Association}, ARMA 14-7014, 12 pp.   


\bibitem{BazCan13}  Ba\v zant, Z.P., and Caner, F.C. 2013, ``Comminution of solids caused by kinetic energy of high shear strain rate, with implications for impact, shock and shale fracturing." {\em P Natl Acad Sci Usa} \textbf{110} (48), 19291--19294.

\bibitem{BazCan14} Ba\v zant, Z.P., and Caner, F.C. 2014, ``Impact comminution of solids due to local kinetic energy of high shear strain rate: I. Continuum theory and turbulence analogy." {\em J Mech Phys Solids} 64,  223--235 (with Corrigendum, Vol. \textbf{67} (2014), p. 14).


























  }
\end{thebibliography}
\end{document}